\documentclass[runningheads,fleqn]{svmult}

 \usepackage{amsmath}
 \usepackage{amssymb}
 \usepackage{makeidx}   
 \usepackage{graphicx}  
 \usepackage{subeqnar}  
 \usepackage{multicol}
 \usepackage{cropmark}
 \usepackage{physmult}
 \usepackage{enumerate}                
 \usepackage[dvips]{color}
 \usepackage{psfrag}        
 \usepackage{epsfig}        
 \usepackage{bm}        

\makeindex

\newcommand{\f}[1]{\mbox{\boldmath$#1$}}
\newcommand{\smallf}[1]{\mbox{\scriptsize\boldmath$#1$}}
\newcommand{\vau}{\mbox{\boldmath$v$}}
\newcommand{\na}{\mbox{\boldmath$\nabla$}}

\begin{document}

\title*{Dynamical Aspects of Analogue Gravity: \\
The Backreaction of Quantum Fluctuations in Dilute
Bose-Einstein Condensates}

\toctitle{Dynamical Aspects of Analogue Gravity:
\protect\newline Quantum Backreaction in Bose-Einstein Condensates}
\titlerunning{Quantum Backreaction in Bose-Einstein Condensates}
\author{Uwe R. Fischer\bigskip\\ 
{\small Eberhard-Karls-Universit\"at T\"ubingen,
Institut f\"ur Theoretische Physik\\
Auf der Morgenstelle 14, D-72076 T\"ubingen, Germany}}
\authorrunning{Uwe R. Fischer}

\maketitle              


\section{Analogue Gravity: An Overview}
\label{Analogue Gravity}



\subsection{The  concept of an effective space-time metric}


Curved space-times \index{space-time!curved} are familiar from Einstein's
theory of gravitation \index{general relativity}
\cite{MTW}, where the metric tensor \index{metric tensor} ${\sf g}_{\mu\nu}$,
describing distances in a curved
space-time with local Lorentz invariance\index{Lorentz invariance!local}, is
determined
by the solution of the Einstein equations\index{Einstein!equations}. A major
problem
for an experimental investigation of the (kinematical as well
as dynamical) properties of curved space-times is that
generating a significant curvature\index{curvature}, equivalent to a
(relatively) small
curvature radius, is a close to impossible undertaking
in manmade laboratories.
For example, the effect of the gravitation of the whole Earth
is to cause
a deviation from flat space-time on this planet's surface
of only the order of $10^{-8}$
(the ratio of Schwarzschild \index{Schwarzschild radius} and Earth radii).
The fact that proper gravitational effects are intrinsically small
is basically due to the smallness of Newton's gravitational constant
\index{gravitational!constant}\index{Newton's gravitational constant}
$G=6.67\times 10^{-11}$\,m$^3$kg$^{-1}$sec$^{-2}$.
Various fundamental classical and quantum
effects in strong gravitational fields
are thus inaccessible for
Earth-based experiments. The realm of strong
gravitational fields \index{gravitational!field} (or, equivalently,
rapidly accelerating a reference frame to simulate gravity
according to the equivalence principle\index{equivalence!principle}), is
therefore difficult to reach.
However, Earth-based gravity experiments are desirable,
because they have the obvious advantage that they
can be prepared and, in particular, repeated under possibly different
conditions at will.

A possible way out of this dilemma, at least  inasmuch the kinematical
properties of curved space-times are concerned,
is the realization of {\em effective} curved space-time geometries
\index{space-time!effective curved} to
mimic the effects of gravity.
Among the most
suitable systems are Bose-Einstein condensates\index{Bose--Einstein
condensate}, i.e., the dilute
matter-wave-coherent gases formed
if cooled to ultralow temperatures\index{temperature!ultra low}, where
the critical temperatures \index{temperature!critical} are of order $T_c\sim
1\,{\rm nK}\cdots 1\, \mu$K;
for reviews of the (relatively) recent status
of this rapidly developing field see \cite{Anglin,LeggettBEC,BECReview}.
In what follows, it will be of some importance that
Bose-Einstein condensates belong to a special class of {quantum}
perfect fluids, so-called superfluids \index{superfluid} \cite{Khalatnikov}.

The curved space-times \index{space-time!curved} we have in mind in
the following are experienced by sound waves \index{sound waves} propagating
on the background of a spatially and temporally inhomogeneous perfect fluid.
Of primary importance is, first of all,
to realize that the identification of sound waves
propagating on an inhomogeneous background, which is
itself determined by a solution of
Euler and continuity equations\index{Euler equation}\index{continuity
equation}, and photons \index{photon}
propagating in a curved space-time, which is determined by a
solution of the Einstein equations\index{Einstein!equations}, is of a {\em
kinematical} nature.
That is, the  space-time metric \index{metric} is fixed externally by a
background
field obeying the laws of hydrodynamics \index{hydrodynamics} (which is
prepared by the
experimentalist), and not self-consistently by a solution of the Einstein
equations.

As a first introductory step to understand the nature of the
kinematical identity,
consider the wave equation for the velocity potential
of the sound field $\phi$, which in a homogeneous medium at rest reads
\begin{equation}
\left[\frac 1{c_s^2} \frac{\partial^2}{\partial t^2}
- \Delta\right]\phi =0, \label{WaveEq}
\end{equation}
where $c_s$ is the sound speed\index{speed!of sound}, which is a constant
in space and time for such a medium at rest.
This equation has Lorentz invariance\index{Lorentz invariance}, that is, if we
replace the speed of light \index{speed!of light} by the speed of sound, it
retains
the form shown above in the new space-time coordinates,
obtained after Lorentz-transforming \index{Lorentz transformation} to a frame
moving at a
constant speed less than the sound speed. Just as the light field
{\em in vacuo} is a proper relativistic field\index{relativistic field}, sound
is a
``relativistic'' field.\footnote{More properly, we should
term this form of Lorentz invariance
{\em pseudorelativistic} invariance. We will
however use for simplicity ``relativistic'' as a generic term
if no confusion can arise therefrom.}
The Lorentz invariance can be made more manifest by writing
equation (\ref{WaveEq}) in the form
$ \Box \phi\equiv
{\sf \eta}^{\mu\nu}\partial_\mu \partial_\nu \phi = 0 $, where
$\eta^{\mu\nu} = $\,diag($1,-1,-1,-1)$ is the (contravariant)
flat space-time metric \index{metric!flat} (we choose throughout the signature
of the metric as specified here), determining the fundamental
light-cone-like structure of Minkowski space \index{Minkowski space}
\cite{minkowski};
we employ the summation convention over equal greek indices $\mu,\nu,\cdots.$
Assuming, then, the sound speed $c_s = c_s ({\f x},t)$
to be local in space and time, and employing the curved space-time version
of the 3+1D Laplacian $\Box$ \cite{MTW}, one can write down the sound wave
equation in a {\em spatially and temporally
inhomogeneous medium} in the generally covariant form
\cite{unruh,visser} 
\begin{equation}
 \frac{1}{\sqrt{\sf -g}} \partial_\mu
(\sqrt{\sf -g}{\sf g}^{\mu\nu}\partial_\nu \phi) =0 .
\label{curvedspaceWaveEq}
\end{equation}
Here, ${\sf g} = {\rm det} [{\sf g}_{\mu\nu}]$ is the
determinant of the (covariant) metric tensor\index{metric tensor}. It is to be
emphasized at this point that, because the space and time
derivatives $\partial_\mu$ are covariantly transforming objects in
(\ref{curvedspaceWaveEq}), the primary object in
the condensed-matter identification of space-time metrics
via the wave equation (\ref{curvedspaceWaveEq}) is the
contravariant metric tensor
${\sf g}^{\mu\nu}$ \cite{GrishaUniverse}. In the condensed-matter
understanding of analogue gravity\index{analogue!gravity}, the quantities 
${\sf g}^{\mu\nu}$
are {\em material-dependent} coefficients. They occur
in a dispersion relation of the
form ${\sf g}^{\mu\nu}k_\mu k_\nu=0$, where $k_\mu = (\omega/c_s, {\f k})$
is the covariant wave vector, with $\hbar {\f k}$ the ordinary spatial
momentum (or quasi-momentum in a crystal).

The contravariant tensor components
${\sf g}^{\mu\nu}$ for a perfect, irrotational liquid \index{fluid!perfect,
irrotational}\index{perfect fluid} turn
out to be  \cite{unruh,visser,trautman}
\begin{equation}
{\sf g}^{\mu\nu}= \frac{1}{A_c c_s^2}  \left( \begin{matrix} 1 & {\f v}
\\ {\f v} & \, -c_s^2 {\f 1} +{\f v} \otimes {\f v}
\end{matrix} \right),\label{gup}
\end{equation}
where $\f 1$ is the unit matrix and $A_c$ a  space and time dependent
function, to be determined from the equations of motion for the
sound field (see below).
Inverting this expression according to
$\textsf{g}^{\beta\nu}\textsf{g}_{\nu \alpha}= \delta^{\beta}{}_\alpha$,
to obtain the covariant metric ${\sf g}_{\mu\nu}$,
the fundamental tensor of distance reads
\begin{equation}
{\sf g}_{\mu\nu}= A_c \left( \begin{matrix} c_s^2-{\f v}^2 & {\f v}
\\ {\f v} & -{\f 1}
\end{matrix} \right), \label{gdown}
\end{equation}
where the line element is $ds^2= {\sf g}_{\mu\nu} dx^\mu dx^\nu$.
This form of the metric has been derived 
by Unruh for an irrotational perfect
fluid described by Euler and continuity equations \cite{unruh};
its properties were later on explored in more detail
in particular by M. Visser \cite{visser}.
We also mention that an earlier derivation of Unruh's form of the metric
exists, from a somewhat different perspective; it was performed by
Trautman \cite{trautman}. The metric belongs to the Painlev\'e-Gullstrand
class of metrics\index{metric!Painlev\'e-Gullstrand class}, historically
introduced in Refs. \cite{PG}.

The conformal factor $A_c$ in (\ref{gdown}) depends on the spatial
dimension of the fluid. It may be unambiguously 
determined by considering the effective
action of the velocity potential fluctuations
above an inhomogeneous background,
identifying this action with the action of a minimally
coupled scalar field \index{scalar!field} in $D+1$-dimensional space-time
\begin{eqnarray}
{\cal A}_{\rm eff}  & = & \int d^{D+1}x 
\frac{1}{2g} \left[ \left(\frac\partial{\partial t} \phi -{\f v}
\cdot \nabla \phi\right)^2 - c_s^2  (\nabla \phi)^2 \right] \nonumber\\
& \equiv & \frac12 \int d^{D+1}x 
\sqrt{-{\sf g}} {\sf g}^{\mu\nu}
\partial_\mu \phi \partial_\nu \phi \,, \label{fischer_action}
\end{eqnarray}
where the prefactor $1/g$ in front of the square brackets in the first line
is identified with the compressibility \index{compressibility} of the
(barotropic)
fluid\index{fluid!barotropic}, $1/g = d(\ln\rho)/dp$, with $p$ the pressure
and $\rho$ the mass density of the fluid; we assume
here and in what follows  that $g$ is a constant independent of space and time
so that $c_s^2 =g\rho$, as valid for a dilute Bose gas (see the following
subsection). Using the above identification, it may
easily be shown that the conformal factor is given by
$A_c = \left({c_s}/{g}\right)^{2/(D-1)}= (\rho/g)^{1/(D-1)})$,
while the square root of the negative determinant is
$\sqrt{-{\sf g}} = c_s (c_s/g)^{D+1/(D-1)}= (\rho^D/g)^{1/(D-1)}$.
The case of one spatial dimension ($D=1$) is special, in that the
conformal invariance \index{conformal!invariance} in two space-time
dimensions implies
that the classical equations of motion are invariant (take the same form)
for any space and time dependent choice of the conformal factor $A_c$,
explaining the singular character of the conformal factor at the special
value $D=1$.

The line element $ds^2 = {\sf g}_{\mu\nu} dx^\mu dx^\nu$
gives us the distances travelled by the phonons \index{phonon}
in an effective space-time world in which
the scalar field $\phi$ ``lives''.
In particular, quasiclassical (large momentum) phonons
\index{phonon!quasiclassical} will follow
{\em light}-like, that is, here, {\em sound}-like geodesics \index{geodesics}
in that
space-time,
according to $ds^2=0$.
Noteworthy is the simple fact that the
constant time slices obtained by setting
$dt=0$ in the line element are conformally flat, i.e.
the quasiparticle world looks on constant time slices
like the ordinary (Newtonian)
lab space, with a Euclidean metric in the case of Cartesian spatial
co-ordinates we display.
All the intrinsic curvature \index{intrinsic curvature} of the effective %
quasiparticle
space-time is therefore
encoded in the metric tensor elements ${\sf g}_{00}$ and
${\sf g}_{0i}$ \cite{Annalspaper}.

\subsection{The metric in Bose-Einstein condensates}

We assumed in Eq.\,(\ref{fischer_action})
that the compressibility \index{compressibility} $1/g$
is a constant.
This entails that the (barotropic) equation of state reads
$p = \frac12 g \rho^2$. We then have, in the microscopic terms
of the interaction between the particles (atoms)
constituting the fluid, 
a contact interaction \index{contact interaction} (pseudo-)potential,
$V({\f x}-{\f x}')= g \delta({\f x}-{\f x}')$.
This is indeed the case for the {dilute} atomic gases \index{gas!dilute
atomic} forming
a Bose-Einstein condensate\index{Bose--Einstein condensate}. Well below the
transition temperature,
they are described, to lowest order in the gas parameter
$(\rho a_s^3)^{1/2}\ll 1$ [where $a_s$ is the $s$-wave scattering
length, assumed positive] by the Gross-Pitaevski\v\i\/
mean-field equation \index{mean-field!equation} for the order parameter
$\Psi \equiv \langle \hat \Psi \rangle$, the
expectation value\footnote{Observe that
$\langle \hat \Psi \rangle\neq 0$ breaks particle number
conservation (the global U(1) invariance).
We will come back to this point in section \ref{Bose--Einstein}, where
we introduce a particle-number-conserving mean-field ansatz
for the full quantum field operator, which has the number-conserving 
property that $\langle \hat \Psi \rangle= 0$, and for which the order
parameter therefore does {not} equal  $\langle \hat \Psi \rangle$.} 
of the quantum field operator $\hat \Psi$ \cite{LeggettBEC}:
\begin{equation}
i\hbar\frac{\partial}{\partial t}\Psi ({\f x},t)
= \left[ -\frac{\hbar^2}{2m} \Delta
+ V_{\rm trap} ({\f x},t) + g |\Psi ({\f x},t)|^2
\right] \Psi ({\f x},t) . \label{GPEq}
\end{equation}
Here, $V_{\rm trap}$ denotes the one-particle trapping potential
\index{trapping potential} 
and the coupling constant $g$ is related to the $s$-wave scattering
length~$a_s$ via ${g=4\pi\hbar^2 a_s/m}$ (in three spatial dimensions).
The Madelung transformation \index{Madelung transformation} decomposing the
complex field \index{complex field} $\Psi$
into modulus and phase reads $\Psi = \sqrt \rho \exp [i\phi]$,
where $\rho$ yields the condensate density and $\phi$ is the
velocity potential. It allows for an interpretation of quantum theory
in terms of hydrodynamics \index{hydrodynamics} \cite{madelung}.
Namely, identifying real and imaginary parts on left- and
right-hand sides of (\ref{GPEq}), respectively,  gives us the two equations
\begin{eqnarray}
-\hbar \frac{\partial}{\partial t} \phi & = &
 \frac12 m {\f v}^2 + V_{\rm trap} + g\rho
-\frac{\hbar^2}{2m} \frac{\Delta {\sqrt\rho}}{\sqrt\rho}
, \label{JAEq}\\
& &
\hspace*{-3.2em}\frac{\partial}{\partial t}\rho + \nabla \cdot(\rho {\f v}) =   
0.
\label{contEq}
\end{eqnarray}
The first of these equations
is the Josephson equation \index{Josephson equation} for the superfluid
\index{superfluid} phase.
This Josephson equation corresponds to the Bernoulli
equation \index{Bernoulli equation} of classical hydrodynamics,
where the usual velocity potential of irrotational hydrodynamics
equals the superfluid phase $\phi$ times $\hbar/m$,
such that ${\f v} = \hbar\nabla\phi/m$.
The latter equation implies that the flow is irrotational save
on singular lines, around which the {\em wave function phase}
$\phi$ is defined only modulo
$2\pi$. Therefore, circulation
is quantized \cite{onsagerstathydro}, and these singular lines
are the center lines of quantized vortices.\index{vortices!quantized}
\index{quantized vortices} 
The usual classical terms in 
Eq.\,(\ref{JAEq})
are augmented by the ``quantum pressure'' \index{quantum!pressure}
$p_Q\equiv -\frac{\hbar^2}{2m} {(\Delta {\sqrt\rho})}/{\sqrt\rho}$. 
The second equation (\ref{contEq})
is the continuity equation for conservation of
particle number, i.e., atom number in the superfluid gas.
The dynamics of the weakly interacting, dilute ensemble of atoms is thus
that of a perfect Euler fluid with quantized circulation of singular
vortex lines. 
This is true except for regions in which the density rapidly varies
and the quantum pressure term $p_Q$ becomes relevant, which happens
on scales of order the coherence length\footnote{Note that
the coherence or ``healing'' length \index{healing length}
is also frequently defined in the literature, see
e.g. \cite{BECReview}, with an additional  factor of $1/\sqrt2$, i.e., as
$\xi_c = \hbar/\sqrt{2gm\rho}$.}
$\xi = \hbar/\sqrt{gm\rho}$
where $\rho$ is a constant (asymptotic) density far away from the
density-depleted (or possibly density-enhanced) region. This is the case
in the density-depleted cores of quantized vortices,\index{quantized vortices}
\index{vortices!quantized} 
or at the low-density boundaries of the system.
The quantum pressure\index{quantum!pressure} is negligible outside these
domains
of rapidly varying and/or low density.
The whole armoury
of space-time metric description of excitations,
 explained in the last section, and
based on the Euler and continuity equations, is then valid
for phonon excitations of a Bose-Einstein condensate, with
the space-time metric (\ref{gdown}), as long as we are outside the
core of quantized vortices and far from the boundaries of the system,
where both the flow is irrotational and the quantum pressure is negligible.

We mention here in passing that the form (\ref{curvedspaceWaveEq})
of the wave equation is valid in quite general physical contexts.
That is, a generally covariant \index{space-time!generally covariant curved}
curved space-time wave equation can be formulated not just
for the velocity perturbation potential in an irrotational
Euler fluid\index{Euler fluid}, for which we have introduced the effective
metric concept.
If the spectrum of excitations (in the local rest frame)
is linear, $\omega = c_{\rm prop} k$, where $c_{\rm prop}$ is the
propagation speed of {\em some} collective excitation\index{collective
excitation}\index{excitation!collective}, the
statement that an effective space-time metric exists is
true, provided we only consider wave perturbations of
a single scalar field $\Phi$
constituting a fixed classical background: More precisely, 
given the generic requirement that the action density \index{action!density} $
\cal L$
is a functional of $\Phi$ and its space-time derivatives
$\partial_\mu \Phi$, i.e. ${\cal L}= {\cal L} [\Phi, \partial_\mu \Phi]$,
the fluctuations \index{fluctuations} $\phi \equiv \delta\Phi$
around some classical background solution $\Phi_0$
of the Euler-Lagrange equations always satisfy a wave equation of
the form of Eq.\,(\ref{curvedspaceWaveEq}), with a possible additional
scalar potential term \cite{BLV} comprising, for example, a mass term for
the scalar field\index{scalar!field}. As a consequence, the
effective metric description also applies, {\it inter alia}, to
the quasiparticle excitations around the gap nodes in the superfluid
$^3\!$He-A \cite{Jacobson},
photon propagation in dielectrics \index{photon}\index{dielectrics}
\cite{dielectric},
and surface waves in shallow water \index{surface waves}\index{waves!surface}
\index{water, shallow} \cite{surfacewaves}.

The analogy between photons propagating on given (curved)  space-time
back\-grounds and phonons in spatially and temporally
inhomogeneous superfluids\index{superfluid} or, more generally,
quantized quasiparticles \index{quasi-particle} with linear quasiparticle
dispersion \index{dispersion!linear} in some background,  allows us to apply
many tools and methods
developed for quantum fields in curved space-times \index{space-time!curved}
\cite{BirrellDavies}. We can therefore conclude that the associated
phenomena occur (provided the fundamental commutation relations of these
quantum fields are fulfilled \cite{BillRalf}). Among these phenomena are
Hawking radiation \index{Hawking radiation} \cite{VisserPRL},
the Gibbons-Hawking effect \index{Gibbons--Hawing effect} in de Sitter
spacetime \index{space-time!de Sitter}\cite{GHEffect},
and cosmological particle production \index{particle production!cosmological}
\index{cosmological!particle production}\cite{CPP,BarceloCPP}.
Furthermore, cosmic inflation \index{cosmic!inflation} and the freezing-in
\index{freezing-in} of quantum
vacuum fluctuations \index{quantum vacuum!fluctuations}\index{vacuum
fluctuations} by the horizon crossing \index{horizon!crossing} of the
corresponding modes
may be simulated \cite{Uhlmann}.
A comprehensive recent review of the subject of analogue gravity
in its broadest sense is given in \cite{BLVReview}.

\subsection{Pseudo-Energy-Momentum Tensor}
An important quantity characterizing the dynamics of the field $\phi$
is the pseudo-energy-mo\-mentum tensor\index{pseudo-energy-momentum tensor},
cf.~\cite{stone}.
Since the equation of motion for the scalar mode,
${\nabla_\mu {\sf g}^{\mu\nu}
\nabla_\nu \phi=0}$, where $\nabla_\mu $ denotes the
space-time-covariant derivative,
is equivalent to covariant energy-momentum balance,
expressed by ${\nabla_\mu T^{\mu\nu}=0}$,
the classical pseudo-energy-momentum tensor reads \cite{stone}
\begin{eqnarray}
\label{pseudo-energy-momentum}
T_{\mu\nu}=(\partial_\mu\phi)(\partial_\nu\phi)-\frac12\,
{\sf g}_{\mu\nu} (\partial_\rho\phi)(\partial_\sigma\phi)
{\sf g}^{\rho\sigma}
\,.
\end{eqnarray}
We already stressed that
the identification of field theoretical effects in curved
space-time by analogy (including the {\em existence} of the
pseudo-energy-mo\-mentum tensor), is of a kinematical nature.
An important question concerns the dynamics, that is, the
{\em backreaction} \index{quantum!backreaction} of the quantum fluctuations
\index{quantum fluctuations} of the scalar field \index{scalar!field} onto
the classical background.\index{classical background}
\index{background!classical} 
Extending the analogy to curved space-times a bit further, one is
tempted to apply the effective-action \index{effective-action method} method
(see, e.g., \cite{BirrellDavies} and \cite{balbinot}).
In the effective-action method, one integrates out fluctuations
of the quantum fields to one-loop order, and then determines
the expectation value of the energy-momentum-tensor by the canonical
identification ${\delta{\cal A}_{\rm eff}}/{\delta {\sf g}^{\mu\nu}}
\equiv  \frac12\,\sqrt{-{\sf g}}\,
\langle\hat T_{\mu\nu}\rangle$.
Since the dependence of the effective action~${\cal A}_{\rm eff}$ on
the degrees of freedom of the background~$\eta$ enters via the
effective metric ${\sf g}^{\mu\nu} ={\sf g}^{\mu\nu}(\eta)$,
one finds the backreaction contribution to the equations of
motion of the $\eta$ by differentiation of the effective action
according to
\begin{eqnarray}
\label{pseudo}
\frac{\delta{\cal A}_{\rm eff}}{\delta\eta}
=
\frac{\delta{\cal A}_{\rm eff}}{\delta {\sf g}^{\mu\nu}}\,
\frac{\delta {\sf g}^{\mu\nu} }{\delta\eta}
=
\frac12\,\sqrt{-{\sf g}}\,
\left\langle\hat T_{\mu\nu}\right\rangle\,
\frac{\delta {\sf g}^{\mu\nu}}{\delta\eta}
\,.
\end{eqnarray}
The precise meaning of the expectation value of the
pseudo-energy-momentum tensor, ${\langle\hat T_{\mu\nu}\rangle}$, is
difficult to grasp in general, due to the non-uniqueness of the vacuum
state in a complicated curved space-time background and the
ultraviolet (UV) renormalization procedure\index{renormalization}.
Adopting a covariant renormalization scheme, the results for
${\langle\hat T_{\mu\nu}\rangle}$ can be classified in terms of
geometrical quantities (cf.~the trace anomaly \cite{BirrellDavies}).

However, in calculating the quantum backreaction using the effective-action
method, one is implicitly making two essential assumptions:
first, that the leading contributions to the backreaction are
completely determined by the effective action in Eq.\,(\ref{fischer_action}),
and, second, that deviations from the low-energy effective action at
high energies do not affect the (renormalized) expectation value of the
pseudo-energy-momentum tensor, ${\langle\hat T_{\mu\nu}\rangle}$.
Since the effective covariance in Eq.\,(\ref{fischer_action}) is only a
low-energy property, the applicability of a covariant renormalization
scheme is not obvious in general.
In the following, we critically examine
the question of whether the two assumptions mentioned above are justified,
e.g., whether ${\langle\hat T_{\mu\nu}\rangle}$ completely determines the
backreaction of the linearized quantum fluctuations.

The remainder of these lecture notes, which is based on the results of the
publication \cite{PRDpaper}, is organized as follows.
In section~\ref{Bose--Einstein}, we give a brief introduction to
Bose-Einstein condensates, and introduce a
particle-number-conserving ansatz for the field operator separated into
condensate, single-particle and multi-particle excitation parts.
In the subsequent section~\ref{backreaction}, based on this ansatz, the
backreaction onto the motion of the fluid
using the full current will be calculated, and it is shown that this yields
a different result than that obtained by the effective-action method.
%
%
%
Afterwards, the failure of the effective-action technique is discussed in
more detail in section~\ref{Effective-Action}.
The cutoff dependence of the pseudo-energy-momentum tensor
(\ref{pseudo-energy-momentum}) is addressed in section~\ref{cutoff}.
As a simple example, we consider the influence of the backreaction
contribution on a static quasi-1D condensate in section~\ref{Simple-Example}.

%

\section{Excitations in Bose-Einstein Condensates}\label{Bose--Einstein}

%

\subsection{Particle-number-conserving mean-field expansion}

In the $s$-wave scattering \index{scattering!s-wave} approximation, a dilute
many-particle
system of interacting bosons is described, on a ``microscopic'' level
corresponding to distance scales much larger than the true range of the
interaction potential,  by the field operator equation of motion in the
Heisenberg picture (we set from now on ${\hbar=m=1}$)
\begin{eqnarray}
\label{Heisenberg}
i\frac{\partial}{\partial t}\hat\Psi=
\left(-\frac{1}{2}\,\na^2+V_{\rm trap}+
g\,\hat\Psi^\dagger\hat\Psi\right)\hat\Psi
\,.
\end{eqnarray}
In the limit of many particles $N\gg1$, in a finite trap at zero
temperature with almost complete condensation, the full field operator
$\hat\Psi$ can be  represented in terms of the
particle-number-conserving mean-field {ansatz} \index{mean--field!ansatz}
\cite{Particle,castin}
\begin{eqnarray}
\label{mean--field}
\hat\Psi=\left(\psi_{\rm c}+\hat\chi+\hat\zeta\right)
{\hat A}{\hat N}^{-1/2}
\,.
\end{eqnarray}
Here, the order parameter~${\psi_{\rm c}={\cal O}(\sqrt{N})}$
\index{parameter!order}
[note that $\psi_c\neq \langle \hat \Psi\rangle$, as opposed
to the $\Psi$ in Eq.\,(\ref{GPEq})]. The one-particle
excitations are denoted ${\hat\chi={\cal O}(N^0)}$, where
{\em one-particle} here means that the Fourier components of $\hat\chi$
are linear superpositions of annihilation and creation operators
of quasiparticles
$\hat a_{\smallf k}$ and $\hat a^\dagger_{\smallf k}$,
cf. Eq.\,(\ref{normal-mode}) below.
The remaining higher-order, multi-particle
corrections are described by ${\hat\zeta={\cal O}(1/\sqrt{N})}$.
The above mean-field {ansatz} can be derived in the dilute-gas limit
by formally setting ${g={\cal O}(1/N)}$ 
\cite{castin,1/N,derivation,meanfield};
we shall use this formal definition of the dilute-gas limit in what
follows.
The dilute-gas limit should be compared and contrasted with the
usual thermodynamic limit\index{thermodynamic limit}, in which the density and
particle interaction
remains constant, while the size of the (trapped) system increases
with $N\rightarrow \infty$, adjusting the harmonic trapping potential
$V_{\rm trap}$
correspondingly (in $D$ spatial dimensions, the thermodynamic limit
corresponds to keeping $N\omega^D$  constant for $N\rightarrow \infty$, where
$\omega$ is the geometric mean of the trapping frequencies \cite{BECReview}).
In the presently used  dilute-gas limit, on the other hand,
the trapping potential remains constant, but the interaction
and the density change.
The advantage of the limit $gN$ constant is that in this limit we have
a well-defined prescription to implement the mean-field approximation,
keeping one power of $g$ for each factor of $N$, cf.~\cite{Lee}.

\subsection{Gross-Pitaevski\v\i\/ and Bogoliubov-de Gennes equations}

Insertion of Eq.\,(\ref{mean--field}) into (\ref{Heisenberg}) yields to
${\cal O}(N)$ 
the Gross-Pitaevski\v\i\/ equation \index{Gross--Pitaevskii equation}
\cite{GP}
for the order parameter~$\psi_{\rm c}$ \index{parameter!order}
\begin{eqnarray}
\label{GP}
i\frac{\partial}{\partial t}\psi_{\rm c}
&=&
\left(
-\frac{1}{2}\,\na^2+V_{\rm trap}+g|\psi_{\rm c}|^2
+2g\left\langle\hat\chi^\dagger\hat\chi\right\rangle
\right)\psi_{\rm c}
+ g\left\langle\hat\chi^2\right\rangle\psi_{\rm c}^*
\,.
\end{eqnarray}
The Bogoliubov-de~Gennes equations \index{Bogoliubov--de~Gennes equations}
\cite{BdG} for
the one-particle fluctuations $\hat\chi$
are obtained to ${\cal O}(N^0)$
\begin{eqnarray}
\label{BdG}
i\frac{\partial}{\partial t}\hat\chi=
\left(
-\frac{1}{2}\,\na^2+V_{\rm trap}+
2g|\psi_{\rm c}|^2\right)\hat\chi
+g\psi^2_{\rm c}\hat\chi^\dagger
\,,
\end{eqnarray}
Finally, the time evolution of the remaining higher-order
corrections in the expansion (\ref{mean--field}),
${\hat\zeta={\cal O}(1/\sqrt{N})}$,
neglecting the ${\cal O}(1/N)$ terms, is given by:
\begin{eqnarray}
\label{xi}
i\frac{\partial}{\partial t}\hat\zeta
&\approx&
\left(
-\frac{1}{2}\,\na^2+V_{\rm trap}+2g|\psi_{\rm c}|^2\right)\hat\zeta
+g\psi^2_{\rm c}\,\hat\zeta^\dagger
\nonumber \\ &&
+ 2g(\hat\chi^\dagger\hat\chi-
\langle\hat\chi^\dagger\hat\chi\rangle)\psi_{\rm c}
+g(\hat\chi^2-\langle\hat\chi^2\rangle)\psi_{\rm c}^*
\,.
\end{eqnarray}
The Gross-Pitaevski\v\i\/ equation in the form (\ref{GP})
ought to be compared with the simple-minded form of
(\ref{GPEq}).
The additional terms ${2g\langle\hat\chi^\dagger\hat\chi\rangle}$ and
${g\langle\hat\chi^2\rangle}$
in the Gross-Pitaevski\v\i\/ equation in the form of Eq.\,(\ref{GP}) above   
ensure that the expectation value of the multi-particle operator,
${\hat\zeta={\cal O}(1/\sqrt{N})}$, vanishes in leading order,
${\langle\hat\zeta\rangle={\cal O}(1/N)}$.
Without these additional terms, the mean-field
expansion~(\ref{mean--field}) would still be valid with
${\hat\zeta={\cal O}(1/\sqrt{N})}$, but without
${\langle\hat\zeta\rangle={\cal O}(1/N)}$.
The proper incorporation of the so-called ``anomalous'' fluctuation average
\index{anomalous fluctuation average}\index{fluctuation average!anomalous}
$\langle\hat\chi^2\rangle$ (the ``normal'' fluctuation average \index{normal
fluctuation average}\index{fluctuation average!normal} is
$\langle\hat\chi^\dagger\hat\chi\rangle$) into
the description of Bose-Einstein
condensates \index{Bose--Einstein condensate} has also been discussed from
various points of view
in \cite{castin,temperature,Yukalova}.

%

\section{Quantum Backreaction}\label{backreaction}

%

\subsection{Calculation of backreaction force from microscopic physics}

The observation that the Gross-Pitaevski\v\i\/ equation~(\ref{GP})
\index{Gross--Pitaevskii equation}
yields an equation correct to leading order ${\cal O}(\sqrt{N})$, using
either $|\psi_{\rm c}|^2$ or
${|\psi_{\rm c}|^2+2\left\langle\hat\chi^\dagger\hat\chi\right\rangle}$
in the first line of (\ref{GP}), hints at the fact that quantum
backreaction \index{quantum!backreaction} effects correspond to
next-to-leading, i.e. quadratic
order terms in the fluctuations \index{quantum fluctuations} and
cannot be derived {\em ab initio} in the above manner without
additional assumptions.
Therefore, we shall employ an alternative method:
In terms of the exact density and current given by
\begin{eqnarray}
\label{exact}
\varrho=\left\langle\hat\Psi^\dagger\hat\Psi\right\rangle
\,,\;
\f{j}
=
\frac{1}{2i}\,
\left\langle\hat\Psi^\dagger\na\hat\Psi
-{\rm H.c.}
\right\rangle
\,,
\end{eqnarray}
the time-evolution is governed by the equation of continuity for
$\varrho$ and an Euler type equation for the current~$\f{j}$.
After insertion of Eq.\,(\ref{Heisenberg}), we find that the equation
of continuity is not modified by the quantum fluctuations but
satisfied exactly (i.e., to all orders in $1/N$ or $\hbar$)
\begin{eqnarray}
\label{continuity-exact}
\frac{\partial}{\partial t}\varrho+\na\cdot\f{j}=0
\,,
\end{eqnarray}
in accordance with the $U(1)$ invariance \index{U(1) invariance} of
the Hamiltonian and the Noether theorem, \index{Noether theorem}  
cf. \cite{balbinot}.
However, if we insert the mean-field expansion~(\ref{mean--field})
\index{mean--field!expansion} and
write the full density as a sum of condensed and non-condensed parts
\begin{eqnarray}
\label{rho-chi}
\varrho=
\varrho_{\rm c}+\left\langle\hat\chi^\dagger\hat\chi\right\rangle
+{\cal O}(1/\sqrt{N})
\,,
\end{eqnarray}
with $\varrho_{\rm c}=|\psi_{\rm c}|^2$, we find that neither part is
conserved separately in general.
Note that this split requires ${\langle\hat\zeta\rangle={\cal O}(1/N)}$, 
i.e., the modifications to the Gross-Pitaevski\v\i\/ equation~(\ref{GP})
discussed above.
Similarly, we may split up the full current
[with ${\varrho_{\rm c}\vau_{\rm c}=\Im(\psi_{\rm c}^*\na\psi_{\rm c})}$]
\begin{eqnarray}
\label{j-chi}
\f{j}=\varrho_{\rm c}\vau_{\rm c}+
\frac{1}{2i}\,
\left\langle\hat\chi^\dagger\na\hat\chi
-{\rm H.c.}
\right\rangle
+{\cal O}(1/\sqrt{N})
\,,
\end{eqnarray}
and introduce an average velocity $\vau$ via ${\f{j}=\varrho\vau}$.
This enables us to {\em unambiguously} define the quantum
backreaction \index{quantum!backreaction} $\f{Q}$ as the following additional
contribution in an
equation of motion for $\f{j}$ analogous to the Euler equation\index{Euler
equation}:
\begin{eqnarray}
\label{Euler}
\frac{\partial}{\partial t}\f{j}=
\f{f}_{\rm cl}(\f{j},\varrho)
+\f{Q}+{\cal O}(1/\sqrt{N})
\,,
\end{eqnarray}
where the classical force density term 
%
\begin{eqnarray}
\f{f}_{\rm cl}(\f{j},\varrho) & = & - \vau\left[\na \cdot (\varrho\vau)\right]
        - \varrho( \vau \cdot\na)\vau
+ \varrho\na\left(\frac{1}{2}
                \frac{\na^2 \sqrt{\varrho}}{\sqrt{\varrho}}
                - V_{\rm trap}- g \varrho \right). 
\end{eqnarray}
Here, ``classical'' means that
the force density contains no explicit quantum fluctuation terms (i.e., only 
those absorbed in the full density and full current), and in addition just 
the ``quantum pressure'', which already occurs on the mean-field level.
Formulation in terms of the conventional Euler equation, i.e.,
using a convective derivative of the velocity defined
by ${\f j}= \rho {\f v}$ giving the acceleration, yields
\begin{eqnarray}
\label{convective}
\left(\frac{\partial}{\partial t}+\f{v}\cdot\na\right)\vau
&=&
-\na\left(
V_{\rm trap}+g\varrho
-\frac{1}{2}\frac{\na^2\sqrt{\varrho}}{\sqrt{\varrho}}
\right)
+\frac{\f{Q}}{\varrho}
+{\cal O}(N^{-3/2})
\,.
\end{eqnarray}
The quantum backreaction force density $\f{Q}$ can now be calculated
by comparing the two equations above and expressing
${\partial\f{j}/\partial t}$ in terms of the field operators via
Eqs.\,(\ref{Heisenberg}) and (\ref{exact})
\begin{eqnarray}
\frac{\partial}{\partial t}\f{j}
&=&
\frac14\left\langle\hat\Psi^\dagger\na^3\hat\Psi-
(\na^2\hat\Psi^\dagger)\na\hat\Psi+{\rm H.c.}\right\rangle
\nonumber \\&&
-\left\langle\hat\Psi^\dagger\hat\Psi\right\rangle\na V_{\rm trap}
-\frac{1}{2g}\na\left\langle g^2(\hat\Psi^\dagger)^2\hat\Psi^2\right\rangle
\,. \label{field-operators}
\end{eqnarray}
After insertion of the mean-field expansion~(\ref{mean--field})
\index{mean--field!expansion},
we obtain the leading contributions in the Thomas-Fermi limit
\index{Thomas--Fermi limit}
\begin{eqnarray}
\label{TF}
\f{Q}
&=&
\na\cdot\left(\vau\otimes\f{j}_\chi+\f{j}_\chi\otimes\vau-
\varrho_\chi\vau\otimes\vau\right)
\nonumber \\
&&-
\frac{1}{2g}\na\left(g^2
\left\langle
2|\psi_{\rm c}|^2\hat\chi^\dagger\hat\chi+
\psi_{\rm c}^2(\hat\chi^\dagger)^2+(\psi_{\rm c}^*)^2\hat\chi^2
\right\rangle
\right)
\nonumber \\&&
-\frac12\,\na\cdot
\left\langle(\na\hat\chi^\dagger)\otimes\na\hat\chi+{\rm H.c.}\right\rangle
\,,
\end{eqnarray}
with ${\varrho_\chi=\left\langle\hat\chi^\dagger\hat\chi\right\rangle}$
and ${\f{j}_\chi=\Im\left\langle\hat\chi^\dagger\na\hat\chi\right\rangle}$.
Under the assumption that the relevant length scales $\lambda$
for variations of, e.g., $\varrho$ and $g$, are much larger than the
healing length \index{healing length} ${\xi=(g\varrho)^{-1/2}}$, we have
neglected
terms containing quantum pressure contributions $\na^2\varrho$ and
$[\na\varrho]^2$, which amounts to the \index{local--density approximation}
Thomas-Fermi or local-density approximation.
These contributions would, in particular, spoil the effective (local)
geometry in Eq.\,(\ref{fischer_action}) (the inclusion of the quantum pressure
to derive a ``nonlocal metric'' \index{metric!nonlocal}\index{nonlocal metric}
has been discussed in \cite{BLVBEC}).

%
%
\subsection{Comparison with effective-action technique}\label{Comparison}
%
%

To compare the expression for the backreaction \index{quantum!backreaction}
force density derived
from the full dynamics of $\hat\Psi$ (\ref{TF})
with the force obtained from Eq.\,(\ref{pseudo}),
we have to identify the scalar field \index{scalar!field}
$\phi$ and the contravariant metric ${\sf g}^{\mu\nu}$.
We already know that phonon modes
with wavelength ${\lambda\gg\xi}$ are described by the action
in Eq.\,(\ref{fischer_action}) in terms of the \index{phase fluctuations}
phase fluctuations~$\phi$
provided that ${\sf g}^{\mu\nu}$ is given by (\ref{gup}), where
$A_c= c_s/g$ and $\sqrt{-{\sf g}}=c_s^3/g^2=\rho^2/c_s$
in three spatial dimensions.
The \index{density!fluctuation} density fluctuations~$\delta\varrho$ are
related to the phase
fluctuations~$\phi$ via
${\delta\varrho=-g^{-1}(\partial/\partial t+\vau\cdot\na)\phi}$.

The variables
$\eta=\left\{\rho_{\rm b}, \phi_{\rm b}\right\}$ or alternatively
$\eta=\left\{\rho_{\rm b}, \nabla\phi_{\rm b}\right\}
=\left\{\rho_{\rm b},{\f v}_{\rm b}\right\}$ in (\ref{pseudo})
are then defined by the expectation values of
density and phase operators according to
\begin{eqnarray}
\label{fischer_split}
\hat\Phi
&=&
\langle\hat\Phi\rangle+\hat\phi=\phi_{\rm b}+\hat\phi
\,,
\nonumber \\
\hat\varrho
&=&
\langle\hat\varrho\rangle+\delta\hat\varrho=\varrho_{\rm b}+\delta\hat\varrho
\,.
\end{eqnarray}
The phase operator can formally
be introduced via the following ansatz for the full
field operator
\begin{eqnarray}
\label{Madelung}
\hat\Psi=e^{i\hat\Phi}\,\sqrt{\hat\varrho}
\,.
\end{eqnarray}
Since $\hat\Phi$ and $\hat\varrho$ do not commute, other forms such as
${\hat\Psi=\sqrt{\hat\varrho}\,e^{i\hat\Phi}}$ would not generate a
self-adjoint $\hat\Phi$ (and simultaneously satisfy
$\hat\Psi^\dagger\hat\Psi=\hat\varrho$).
Note that, in contrast to the full density which is a well-defined and
measurable quantity, the velocity potential, $\hat\Phi$, is not
\cite{Froehlich}. This can be seen as follows. 
The commutator between density and phase operators
\begin{equation}
[\hat\varrho({\f r}),\hat\Phi({\f r}')] = i\delta ({\f r}-{\f r}'),
\label{canComm} 
\end{equation}
yields, if one takes, on both sides,  matrix elements in the number basis 
of the space integral over $\bm r$ for a given volume $\cal V$ in which 
$\bm r'$ is situated,    
\begin{equation} 
\label{intComm} 
(N-N')\langle N|\hat\Phi({\f  r}')|N'\rangle = i\delta_{NN'}, 
\end{equation}  
where $N$ and $N'$ are two possible values for the number of
particles in $\cal V$.  
This is an inconsistent relation for $N,N'$ positive semidefinite 
and discrete (which the very existence of particles requires), 
most obviously for $N=N'$. 
The commutator therefore makes sense only if it is understood to be
effectively coarse-grained over a sub-volume $\cal V$ 
with large enough number of particles $N\gg 1$, such that
the inconsistency inherent in (\ref{intComm})  
becomes asymptotically irrelevant.
It cannot be defined consistently locally, i.e., in arbitrarily small
volumina, where there is just one particle or even none, or when the number
fluctuations in larger volumina $\cal V$ are large, such that the probability 
to have a very small number of particles in $\cal V$ is not negligible 
\cite{Castin}.

The action in terms of the total density $\varrho$
and the variable $\Phi$ reads (neglecting the quantum pressure term, i.e.,
in the Thomas-Fermi limit)
\begin{eqnarray}
{\cal L}
=
-\varrho\left(\frac{\partial}{\partial t}\Phi+\frac12(\na\Phi)^2\right)
-\epsilon[\varrho] -V_{\rm trap}\varrho
\,,\label{nonfundaction}
\end{eqnarray}
with $\epsilon[\varrho]$ denoting the internal energy density.
The quantum corrections to the Bernoulli equation up to second order
in the fluctuations, using the effective-action method in Eq.\,(\ref{pseudo})
are then incorporated by writing (the background density here
equals the full density, $\varrho=\varrho_{\rm b}$)
\begin{eqnarray}
\frac{\partial}{\partial t} \phi_{\rm b}
+ \frac12 (\nabla \phi_{\rm b})^2 + h[\varrho]
-\frac{\delta {\cal A}_{\rm eff}}{\delta \varrho}=0\,,
\label{BernoulliI}
\end{eqnarray}
where  $h[\varrho] = d\epsilon/d\varrho+V_{\rm trap}$.
We obtain, using (\ref{fischer_action}), 
%
\begin{eqnarray}
\label{corrections}
\frac{\delta{\cal A}_{\rm eff}}{\delta\varrho}
=\frac{\delta{\cal A}_{\rm eff}}{\delta{\sf g}^{\mu\nu}}
\frac{\delta {\sf g}^{\mu\nu}}{\delta\varrho}
= \frac{\sqrt{-{\sf g}}}{2}
{\langle\hat T_{\mu\nu}\rangle} 
\frac{\delta {\sf g}^{\mu\nu}}{\delta\varrho}
= -\frac12\langle(\na\hat\phi)^2\rangle
\,.
\end{eqnarray}
%
Clearly, taking the gradient of
this result we obtain a backreaction \index{quantum!backreaction} force which
markedly
differs from the expression~(\ref{TF}) derived
in the previous subsection.
Moreover, it turns out that the backreaction force density in
Eq.\,(\ref{TF}) contains contributions which
are not part of the expectation value of the
pseudo-energy-momentum tensor\index{pseudo-energy-momentum tensor}, ${\langle
\hat T_{\mu\nu}\rangle}$.
For example, the phonon density $\varrho_\chi$ contains
${\langle(\delta\hat\varrho)^2\rangle_{\rm ren}}$
(where $\langle\dots\rangle_{\rm ren}$ means that the divergent
c-number ${\hat\chi\hat\chi^\dagger-\hat\chi^\dagger\hat\chi=\delta(0)}$
has been subtracted already) which is part of 
${\langle\hat T_{\mu\nu}\rangle_{\rm ren}}$, but also
${\langle\hat\phi^2\rangle_{\rm ren}}$ which is not.
[Note that ${\langle\hat\phi^2\rangle_{\rm ren}}$ cannot be cancelled
by the other contributions.] 
%
The expression
${\langle(\na\hat\chi^\dagger)\otimes\na\hat\chi+{\rm H.c.}\rangle}$
in the last line of Eq.\,(\ref{TF}) contains
${\langle\na\hat\phi\otimes\na\hat\phi\rangle_{\rm ren}}$
which does occur in ${\langle\hat T_{\mu\nu}\rangle_{\rm ren}}$, but also
${\langle\na\delta\hat\varrho\otimes\na\delta\hat\varrho\rangle_{\rm ren}}$,
which does not.
One could argue that the latter term ought to be neglected in the
Thomas-Fermi or
local-density approximation since it is on the same footing as the
quantum pressure contributions containing
$\na^2\varrho$ and  $[\na\varrho]^2$ [which have been  neglected
in (\ref{TF})], but it turns out that this expectation value yields
cutoff dependent contributions of the same order of magnitude as
the other terms, see section~\ref{cutoff} below.

%
%
\section{Failure of Effective-Action Technique}\label{Effective-Action}
%
%

After having demonstrated the failure of the effective-action method
\index{effective-action method} for
deducing the quantum backreaction\index{quantum!backreaction}, let us study
the reasons for this
failure in more detail.
The full action governing the dynamics of the fundamental field
$\Psi$ reads
\begin{eqnarray}
{\cal L}^\Psi
=
i\Psi^*\frac{\partial}{\partial t}\Psi
-
\Psi^*\left(-\frac{1}{2}\,\na^2+V_{\rm trap}+\frac{g}{2}\,\Psi^*\Psi\right)
\Psi
\,.
\end{eqnarray}
Linearization according to
$\Psi=\psi_{\rm c}+\chi$ yields the effective second-order
action generating the Bogoliubov-de~Gennes equations~(\ref{BdG})
\index{Bogoliubov--de~Gennes equations}
\begin{eqnarray}
{\cal L}_{\rm eff}^\chi
&=&
i\chi^*\frac{\partial}{\partial t}\chi
-
\chi^*\left(-\frac12 \na^2 +V_{\rm trap}+2g |\psi^2_{\rm c}|\right)\chi
-\frac12\left[g(\psi^*_{\rm c})^2\chi^2+{\rm H.c.}\right]
\,. \nonumber \\ &&
\end{eqnarray}
%
If we start with the action (\ref{nonfundaction})
in terms of the total density $\varrho$
and the nonfundamental variable $\Phi$,
the quantum corrections to the equation of continuity \index{equation of
continuity!quantum corrections}
$\delta{\cal A}_{\rm eff}/\delta\phi_{\rm b}$ are reproduced correctly
but the derived quantum backreaction contribution to the Bernoulli
\index{Bernoulli equation!quantum corrections}
equation, $\delta{\cal A}_{\rm eff}/\delta\varrho_{\rm b}$, and therefore
the backreaction force, is wrong.

One now is led to the question why the effective-action method
works for the fundamental field
$\Psi$ and gives the correct expression for the backreaction force,
but fails for the non-fundamental variable $\Phi$.
The quantized fundamental field $\hat\Psi$ satisfies the equation of
motion (\ref{Heisenberg}) as derived from the above action and possesses
a well-defined linearization via the mean-field expansion
(\ref{mean--field}).
One of the main assumptions of the effective-action method is a similar
procedure for the variable $\Phi$, i.e., the existence of a well-defined
and linearizable full quantum operator $\hat\Phi$ satisfying the
quantum Bernoulli equation \index{Bernoulli equation!quantum} (for large
length scales), cf. Eq.\,(\ref{BernoulliI})
\begin{eqnarray}
\frac{\partial}{\partial t}\hat\Phi+\frac12(\na\hat\Phi)^2
+h[\hat\varrho]
\stackrel{?}{=}0
\,.\label{Bguess}
\end{eqnarray}
%
%
The problem is that the commutator of $\hat\varrho$ and $\hat\Phi$ at the
same position diverges, cf. Eq.\,(\ref{canComm}), 
and hence the quantum Madelung ansatz in
Eq.\,(\ref{Madelung}) is singular.
As a result, the above quantum Bernoulli equation is not well-defined
(in contrast to the equation of continuity), i.e., insertion of the
quantum Madelung ansatz in Eq.\,(\ref{Madelung}) into Eq.\,(\ref{Heisenberg})
generates divergences \cite{Froehlich}.

In order to study these divergences by means of a simple example,
let us consider a generalized Bose-Hubbard Hamiltonian \index{Bose--Hubbard
model} \cite{Jaksch},
which considers bosons sitting on a lattice with sites $i$, which can hop
between nearest neighbor sites and interact if at the same site: 
\begin{eqnarray}
\hat H
=
-\frac\alpha 2\sum\limits_{<ij>}(\hat\Psi^\dagger_i\hat\Psi_j+{\rm H.c.})
+
\sum\limits_i\left(\beta_i\hat n_i+\frac\gamma2\hat n_i^2\right)
\,,\label{BH}
\end{eqnarray}
where $\hat\Psi_i$ is the annihilation
operator for bosons at a given lattice site $i$, and
$<ij>$ denotes summation over nearest neighbors;
$\hat n_i=\hat\Psi^\dagger_i\hat\Psi_i$ is
the so-called filling factor (operator),
equal to the number of bosons at the lattice site $i$. The
quantities $\beta_i$ multiplying the filling factor depend on the site index.
%
In the continuum limit, the lattice Hamiltonian  (\ref{BH})
generates a version of Eq.\,(\ref{Heisenberg}).
Setting $a^{D/2} \hat \Psi ({\f x}_i) = \hat \Psi_i$,
where the   $\hat \Psi ({\f x}_i)$ are the continuum field operators and
$a$ is the lattice spacing taken to zero (we consider for simplicity a simple
cubic lattice in $D$ spatial dimensions), the {\em effective mass}
\index{effective mass!Bose--Hubbard model} is given by
$1/m^* = \alpha a^2$: The bosons moving through the lattice
obviously become the heavier the smaller
the hopping amplitude $\alpha$ becomes at given $a$. 
The coupling constant is determined by
$g= \gamma a^D$, and the trap potential  is
governed by $V_{\rm trap}({\f x})
= g/2 + \beta({\f x})-\alpha$.

On the other hand, inserting the quantum Madelung ansatz employing a phase
operator, Eq.\,(\ref{Madelung}) in its
lattice version,
the problem of operator ordering arises and the (for the Bernoulli
equation) relevant kinetic energy term reads
\begin{eqnarray}
\hat H_\Phi
=
\frac14\sum\limits_i\sqrt{\hat n_i(\hat n_i+1)}\;(\na\hat\Phi)^2_i+{\rm H.c.}
\,,
\end{eqnarray}
with the replacement $\hat n_i+1$ instead of $\hat n_i$ being one
effect of the non-commu\-tativity.
In the superfluid \index{superfluid} phase with large filling $n\gg1$, we
therefore obtain the following
leading correction to the equation of motion
\begin{eqnarray}
\frac{\partial}{\partial t}\hat\Phi+\frac12(\na\hat\Phi)^2
+h[\hat\varrho]
+\frac{1}{\hat n}\,\frac{(\na\hat\Phi)^2}{16}\,\frac{1}{\hat n}
=
{\cal O}\left(\frac{1}{n^3}\right)
\,.
\end{eqnarray}
The Bernoulli equation in its quantum version thus
receives corrections depending on microscopic details
like the filling of a particular site and is not of the (conjectured)
form (\ref{Bguess}).

By means of this simple example, we already see that the various limiting
procedures such as
the quantization and subsequent mean-field expansion,
the variable transformation $(\Psi^*,\Psi)\leftrightarrow(\varrho,\Phi)$,
and the linearization for small fluctuations,
as well as continuum limit
do not commute in general -- which explains the failure of the
effective-action method for deducing the quantum backreaction.
The variable transformation $(\Psi^*,\Psi)\leftrightarrow(\varrho,\Phi)$
is applicable to the zeroth-order equations of motion for the classical
background as well as to the first-order dynamics of the linearized
fluctuations -- but the quantum backreaction is a second-order effect,
where the aforementioned difficulties, such as the question of the choice
of fundamental variables and their operator ordering, arise.

%
%
\section{Cutoff Dependence of Effective Action}\label{cutoff}
%
%

Another critical issue for the applicability of the effective-action method
\index{effective-action method!cutoff dependence}
is the UV divergence \index{UV divergence} of ${\langle\hat T_{\mu\nu}\rangle}
$.
Extrapolating the low-energy effective action in
Eq.\,(\ref{fischer_action}) to large momenta $k$, the expectation values
${\langle\delta\hat\varrho^2\rangle}$ and
${\langle\hat\phi^2\rangle}$ entering $\varrho_\chi$ would diverge.
For Bose-Einstein condensates\index{Bose--Einstein condensate}, we may infer
the deviations from
Eq.\,(\ref{fischer_action}) at large $k$ from the Bogoliubov-de~Gennes
equations~(\ref{BdG}).
Assuming a static and homogeneous background (which should be a good
approximation at large~$k$), a normal-mode expansion yields a Bogoliubov
transformation between the bare bosonic operators $\hat\chi_{\smallf k}$
and the
quasiparticle operators $\hat a_{\smallf k}$, $\hat a_{\smallf k}^\dagger$:
\begin{eqnarray}
\label{normal-mode}
\hat\chi_{\smallf k}
=
\sqrt{\frac{{\f k}^2}{2\omega_{\smallf k}}}
\left[
\left(\frac{\omega_{\smallf k}}{{\f k}^2}-\frac12\right)
\hat a_{\smallf k}^\dagger
+
\left(\frac{\omega_{\smallf k}}{{\f k}^2}+\frac12\right)
\hat a_{\smallf k}
\right]
\,,
\end{eqnarray}
where
the frequency~$\omega_{\smallf k}$ is determined
by the Bogoliubov dispersion relation \index{dispersion relation!Bogoliubov}
for the dilute Bose gas,
${\omega_{\smallf k}^2=g\varrho\,{\f k}^2+{\f k}^4/4}$.
The above form of the Bogoliubov transformation
results, after inversion, in the usual phonon quasiparticle
operators at low momenta, and gives
$\hat \chi_{\smallf k} = \hat a_{\smallf k}$ at ${\f k}\rightarrow \infty$,
i.e.,
the quasiparticles and the bare bosons become, as required,
identical at large momenta.

Using a linear dispersion \index{dispersion!linear}
${\omega_{\smallf k}^2\propto{\f k}^2}$ instead of the full Bogoliubov
dispersion, expectation values such as
${\langle\hat\chi^\dagger\hat\chi\rangle}$
would be UV divergent, but the correct dispersion relation implies
${\hat\chi_{\smallf k}\sim
\hat a_{\smallf k}^\dagger\,g\varrho/{\f k}^2+
\hat a_{\smallf k}}$ for large ${\f k}^2$, and hence
${\langle\hat\chi^\dagger\hat\chi\rangle}$ is UV finite in three and
lower spatial dimensions.
Thus the healing length $\xi$ acts as an effective UV cutoff\index{healing
length!UV cutoff},
$k^{\rm cut}_\xi\equiv 1/\xi$.

Unfortunately, the quadratic decrease for large~$k$ in
Eq.\,(\ref{normal-mode}), giving asymptotically
${\hat\chi_{\smallf k}\sim
\hat a_{\smallf k}^\dagger\,g\varrho/{\f k}^2+
\hat a_{\smallf k}}$,
is not sufficient for rendering the other expectation values
(i.e., apart from $\varrho_\chi$ and $\f{j}_\chi$)
in Eq.\,(\ref{TF}) UV finite in three spatial dimensions.
This UV divergence indicates a failure of the $s$-wave
pseudo-potential ${g\delta^3(\f{r}-\f{r'})}$ in Eq.\,(\ref{Heisenberg})
at large wavenumbers~$k$ and can be eliminated by replacing
${g\delta^3(\f{r}-\f{r'})}$ by a more appropriate two-particle
interaction potential ${V_{\rm int}(\f{r}-\f{r'})}$, see \cite{Lee}.
Introducing another UV cutoff wavenumber~$k^{\rm cut}_s$ related to
the breakdown of the $s$-wave pseudo-potential, we obtain
${\langle(\na\hat\chi^\dagger)\otimes\na\hat\chi+{\rm H.c.}\rangle
\sim g^2\varrho^2\,k^{\rm cut}_s}$
and ${\langle\hat\chi^2\rangle\sim g\varrho\,k^{\rm cut}_s}$.

In summary, there are two different cutoff wavenumbers:
The first one, $k^{\rm cut}_\xi$, is associated to the breakdown of the
effective Lorentz invariance (change of dispersion relation from linear
to quadratic) and renders some -- but not all -- of the naively
divergent expectation values finite.
The second wavenumber, $k^{\rm cut}_s$, describes the cutoff for all
(remaining) UV divergences.
In dilute Bose-Einstein condensates, these two scales are vastly
different by definition. Because the system is dilute, the inverse
range of the true potential $k^{\rm cut}_s\equiv 1/r_0$, must
be much larger than the inverse healing length.
Thus the following condition of scale separation must hold:
\begin{eqnarray}
k^{\rm cut}_{\rm UV}
=
k^{\rm cut}_s
\gg
k^{\rm cut}_\xi
=
k^{\rm cut}_{\rm Lorentz}
\,.
\end{eqnarray}
In terms of the length scales governing the system,
using that $k_\xi^{\rm cut} = \sqrt{4\pi a_s\rho}$,  in the dilute gas
we must have the following condition fulfilled,
$4\pi a_s r_0^2 \ll d^3$, where
$d=\varrho^{-1/3}$ is the interparticle separation.
Note that the opposite scale separation relation,
$k^{\rm cut}_{\rm Lorentz} \gg k^{\rm cut}_{\rm UV}$
is very unnatural in the sense that every quantum field theory which
has the usual properties such as locality and Lorentz invariance 
must have UV divergences (e.g., in the two-point function).

The renormalization of the cutoff-dependent terms is different for the
two cases:
The $k^{\rm cut}_s$-contributions can be absorbed by a
$\varrho$-independent renormalization of the coupling $g$
\cite{meanfield,Lee}, whereas the
$k^{\rm cut}_\xi$-contributions depend on the density in a nontrivial
way and thus lead to a quantum
renormalization of the effective equation of state.
We supply an example for this
renormalization in the section to follow. 

%
%
\section{Static Example for the Backreaction Force}\label{Simple-Example}
%
%

In order to provide an explicit example for the quantum backreaction
\index{quantum!backreaction}
term in Eq.\,(\ref{TF}), without facing the above discussed
UV problem, let us consider
a quasi-one-dimensional (quasi-1D) condensate \index{Bose--Einstein
condensate!quasi-1D}
\cite{1D,Goerlitz}, where all the involved
quantities are UV finite. In a quasi-1D condensate the   
perpendicular harmonic trapping $\omega_\perp$ is much larger than the axial
trapping $\omega_z$ such that the condensate assumes the shape of a strongly
elongated cigar with all atomic motion in the perpendicular direction frozen
out, $\omega_\perp$ being much larger than the mean energy per particle.
%

In accordance with general considerations \cite{Uncertainty},
the phonon density $\varrho_\chi$ is infrared (IR) divergent
in one spatial dimension, therefore inducing finite-size effects, i.e.,
a dependence of the various quantities of interest on the system size.
Nevertheless, in certain situations, we are able to derive a closed
local expression for the quantum backreaction term $Q$:
Let us assume a completely static condensate ${\vau=0}$
in effectively one spatial dimension, still allowing for a spatially
varying density $\varrho$ and possibly also coupling $g$.
Furthermore, since we require that
spatial variations of $\varrho$ and $g$ occur
on length scales $\lambda$ much larger than the healing length
(Thomas-Fermi approximation),
we keep only the leading terms in $\xi/\lambda\ll 1$, i.e., the
variations of $\varrho$ and $g$ will be neglected in the calculation of
the expectation values. 
In this case, the quantum backreaction term $Q$
simplifies considerably and yields
(in effectively one spatial dimension, where ${g\equiv g_{\rm 1D}}$
and ${\varrho\equiv\varrho_{\rm 1D}}$ now both refer to the 1D
quantities) 
\begin{eqnarray}
\label{1+1}
Q
&=&
-
\nabla\left\langle(\nabla\hat\chi^\dagger)\nabla\hat\chi\right\rangle
-
\frac{1}{2g}\nabla\left(g^2\varrho\left\langle
2\hat\chi^\dagger\hat\chi+(\hat\chi^\dagger)^2+\hat\chi^2
\right\rangle
\right)
\nonumber \\
&=&
-\nabla\left(\frac{1}{3\pi}\,(g\varrho)^{3/2}\right)
+\frac{1}{2\pi g}\nabla\left(g^{5/2}\varrho^{3/2}\right)
+{\cal O}(\xi^2/\lambda^2)
\nonumber \\
&=&
\frac{\varrho}{2\pi}\,\nabla\sqrt{g^3\varrho}
+{\cal O}(\xi^2/\lambda^2)
\,.
\end{eqnarray}
It turns out that the IR divergences \index{IR divergence} of
${2\langle\hat\chi^\dagger\hat\chi\rangle}$ and
${\langle(\hat\chi^\dagger)^2+\hat\chi^2\rangle}$ cancel each other
such that the resulting expression is not only UV but also IR finite.
Note that the sign of~$Q$ is positive and hence opposite to the contribution
of the pure phonon density ${\langle\hat\chi^\dagger\hat\chi\rangle}$,
which again illustrates the importance of the ``anomalous'' term
${\langle(\hat\chi^\dagger)^2+\hat\chi^2\rangle}$.

A possible experimental signature of the quantum
backreaction term $Q$ calculated above, is the change
incurred on the static Thomas-Fermi solution of the
Euler equation~(\ref{convective}) for the density distribution
(cf.~\cite{GP,Lee})
\begin{eqnarray}
\label{exp}
\varrho_{\rm 1D}
=
\frac{\mu-V_{\rm trap}}{g_{\rm 1D}}+\frac{\sqrt{\mu-V_{\rm trap}}}{2\pi}
+{\cal O}(1/\sqrt{N})
\,,
\end{eqnarray}
with $\mu$ denoting the (constant) chemical potential.
The classical [${\cal O}(N)$] density profile
${\varrho_{\rm cl}=(\mu-V_{\rm trap})/g_{\rm 1D}}$ acquires nontrivial quantum
[${\cal O}(N^0)$] corrections ${\varrho_Q=\sqrt{\mu-V_{\rm trap}}/2\pi}$, 
where the small parameter is the ratio of the interparticle distance
${1/\varrho={\cal O}(1/N)}$ over the healing
length~${\xi={\cal O}(N^0)}$.
Note that the quantum backreaction term
${\varrho_Q}$ in the above split
${\varrho=\varrho_{\rm cl}+\varrho_Q}$
should neither be confused with the phonon density $\varrho_\chi$ in
${\varrho=\varrho_{\rm c}+\varrho_\chi}$ (remember that $\varrho_\chi$ is IR
divergent and hence contains finite-size effects) nor
with the quantum pressure contribution $\propto\na^2\sqrt{\varrho}$ in
the Euler type Eq.\,(\ref{convective}).

Evaluating the change $\Delta R$ of the Thomas-Fermi size 
(half the full length), where ${\mu=V_{\rm trap}}$, of a quasi-1D
Bose-Einstein condensate induced by backreaction,
from Eq.\,(\ref{exp}) we get
${\Delta R=-2^{-5/2}(\omega_\perp/\omega_z)a_s}$.
Here, the quasi-1D coupling constant $g_{\rm 1D}$ is related to the 3D
$s$-wave scattering length $a_s$ and the perpendicular harmonic
trapping $\omega_\perp$ by ${g_{\rm 1D}=2a_s\omega_\perp}$
(provided $a_s \ll a_\perp = 1/\sqrt{\omega_\perp}$ \cite{1D}).
In units of the classical size
$R_{\rm cl}=(3a_sN\omega_\perp/\omega_z^2)^{1/3}$,
we have     
\begin{equation}
\label{radius}
\frac{\Delta R}{R_{\rm cl}}
=
-\frac{1}{4\sqrt2}
\left(\frac{1}{3N}\right)^{1/3}
\left(\frac{\omega_\perp}{\omega_z}
\frac{a_s}{a_z}\right)^{2/3},
\end{equation}   
where $a_z=1/\sqrt{\omega_z}$ describes the longitudinal harmonic
trapping length.
In quasi-1D condensates, backreaction thus leads to a
{\em shrinking} of the cloud relative to the classical expectation --
whereas in three spatial dimensions we have the opposite effect
\cite{BECReview,Lee}. In one dimension, we thus obtain a {\em softening}
of the quantum renormalized equation of state of the gas; conversely,
in three spatial dimensions the effective equation of state becomes
{\em stiffer} due to quantum fluctuations.

For reasonably realistic experimental parameters, the effect of
quantum backreaction on the equation of state should be measurable; for
${N\simeq 10^3}$, ${\omega_\perp/\omega_z\simeq 10^3}$,
and ${a_s/a_z\simeq 10^{-3}}$,
we obtain ${|\Delta R/{R_{\rm cl}}|\simeq1\%}$.

%
%
\section{Conclusion}\label{Conclusion}
%
%
By explicit analysis of the analytically tractable case of a dilute Bose
gas in the mean-field approximation, we have demonstrated the following. 
Even given that the explicit form of the quantum backreaction
\index{quantum!backreaction} terms depends
on the definition of the classical background, the effective-action method
does not yield the correct result in the general case (which is,
in particular, independent of the choice of variables). 
The knowledge of the classical (macroscopic) equation of motion -- such as
the Bernoulli equation -- may be sufficient for deriving the first-order
dynamics of the linearized quantum fluctuations (phonons), but the quantum
backreaction as a second-order effect cannot be obtained without further
knowledge of the microscopic structure (which reflects itself, for
example, in the operator ordering imposed).
It is tempting to compare these findings to gravity, where we also know the
classical equations of motion only
\begin{eqnarray}
R_{\mu\nu}-\frac12\,{\sf g}_{\mu\nu}\,R= \frac{8\pi G}{c^4} \,T_{\mu\nu}
\,,
\end{eqnarray}
which -- in analogy to the Bernoulli equation -- might yield the correct
first-order equations of motion for the linearized gravitons, but perhaps
not their (second-order) quantum backreaction\index{quantum!backreaction for
gravity}.
Another potentially interesting point of comparison is the existence of two
different high-energy scales -- one associated to the breakdown of Lorentz
invariance \index{Lorentz invariance!breakdown} $k^{\rm cut}_\xi=k^{\rm
cut}_{\rm Lorentz}$ and the other one,
$k^{\rm cut}_{\rm UV}=k^{\rm cut}_s$,  to the UV cutoff introduced by the
true interaction potential range. The question then poses itself
whether one of the two cutoff scales (or in some sense both of them)
correspond to the Planck scale \index{Planck!scale} in gravity.

The dominant ${{\cal O}(\xi/\lambda)}$ quantum backreaction contributions
like those in Eq.\,(\ref{exp}) depend on the healing length as the lower
UV cutoff and hence cannot be derived from the low-energy effective action
in Eq.\,(\ref{fischer_action}) using a covariant (i.e., cutoff independent) 
regularization scheme, which does not take into account details of
microscopic physics (represented, for example, in the
quasiparticle dispersion relation).
Note that the leading ${{\cal O}(\xi/\lambda)}$ quantum correction to the
pressure \index{pressure!quantum corrections} could be identified with a
cosmological term\index{cosmological!term},
${\langle\hat T_{\mu\nu}\rangle=\Lambda\,{\sf g}_{\mu\nu}}$
in Eq.\,(\ref{pseudo}), provided that the cosmological ``constant''
$\Lambda$ is not constant but depends on $g$ and $\varrho$.
Note that
in general relativity\index{general relativity}, the Einstein equations demand
$\Lambda$ to be
constant, due to the equivalence principle and the resulting
requirement that the metric be parallel-transported,
${\nabla^\mu {\sf g}_{\mu\nu}=0}$.

As became evident, 
the knowledge of the expectation value of the pseudo-energy-momentum
\index{pseudo-energy-momentum tensor}
tensor ${\langle\hat T_{\mu\nu}\rangle}$ is not sufficient for
determining the quantum backreaction effects in general.
Even though ${\langle\hat T_{\mu\nu}\rangle}$ is a useful concept
for describing the phonon kinematics (at low energies),
we have seen that it does not represent the full dynamics
of the fluid dynamical variables defined in terms
of the fundamental quantum field $\hat\Psi$.
Related limitations of the {\em classical} pseudo-energy-momentum tensor,
in particular the background choice dependence of the 
description of the second-order effect of the exchange of energy and momentum
between excitations and that background, and the resulting
form of the conservation laws, have been discussed in \cite{stone}. 

In general, the quantum backreaction corrections to the Euler \index{Euler
equation!quantum corrections}
equation in Eq.\,(\ref{convective}) cannot be represented as the
gradient of some local potential, cf.~Eq.\,(\ref{TF}).
Hence they may effectively generate vorticity \index{vorticity} and might
serve as the
seeds for vortex nucleation from the vortex vacuum.
%

In contrast to the three-dimensional case (see, e.g., \cite{GP,Lee}),
the quantum backreaction corrections given by Eq.\,(\ref{1+1})
{\em diminish} the pressure in condensates that can be described
by Eq.\,(\ref{Heisenberg}) in one spatial dimension (quasi-1D case).
This is a direct consequence of the so-called ``anomalous'' term
${\langle(\hat\chi^\dagger)^2+\hat\chi^2\rangle}$ in Eq.\,(\ref{1+1}),
which -- together with the cancellation of the IR divergence --
clearly demonstrates that it cannot be neglected in general.
We emphasize that even though Eqs.\,(\ref{1+1})-(\ref{radius}) describe
the {\em static} quantum backreaction corrections to the ground
state, which can be calculated by an alternative method~\cite{Lee} as
well, the expression in  Eq.\,(\ref{TF}) is valid for more general
dynamical situations, such as rapidly
expanding condensates.  Quantum backreaction
can thus generally {not}
be incorporated by rewriting the Euler equation
in terms of a renormalized chemical potential.
While the static quantum backreaction corrections to the ground state can
be absorbed by a redefinition of the chemical potential
$\mu(\varrho)$ determining a quantum renormalized
(barotropic) equation of state $p(\varrho)$,
this is not possible for the other terms in Eq.\,(\ref{TF}), like 
the quantum friction-type terms depending on $\f{j}\otimes\vau$.

We have derived, from the microscopic physics of dilute Bose-Einstein
condensates, the backreaction of quantum fluctuations onto the motion
of the full fluid and found a quantum backreaction
force that is potentially experimentally observable in existing
condensates.
We observed a failure of the effective-action technique
to fully describe the backreaction force in Eq.\,(\ref{TF}),
and a cutoff dependence of backreaction
due to the breakdown of covariance at high energies.
Whether similar problems, in particular the question of the
correct choice of the fundamental variables and the related operator
ordering issues, beset the formulation of a theory of
``real'' (quantum) gravity remains an
interesting open question.


\bigskip




\begin{thebibliography}{8.}

\addcontentsline{toc}{section}{References}


\bibitem{MTW} C.\,W. Misner, K.\,S. Thorne, and J.\,A. 
Wheeler, {\em Gravitation} (Freeman, 1973).
\bibitem{Anglin} J.\,R. Anglin and W. Ketterle,
``{Bose-Einstein condensation of atomic gases}'',
Nature {\bf 416}, 211 (2002).
\bibitem{LeggettBEC} A.\,J.~Leggett,
``{Bose-Einstein condensation in the alkali gases:
Some fundamental concepts}'', Rev. Mod. Phys. {\bf 73}, 307 (2001).
\bibitem{BECReview}
F.~Dalfovo, S.~Giorgini, L.\,P.~Pitaevski\v\i\/, and S.~Stringari,
``{Theory of Bose-Einstein condensation in trapped gases}'',
Rev.\ Mod.\ Phys.\ {\bf 71}, 463 (1999).
\bibitem{Khalatnikov} I.\,M. Khalatnikov, {\em An Introduction to the
Theory of Superfluidity} (Addison Wesley, Reading, MA, 1965).
\bibitem{minkowski} H. Minkowski, ``{Raum und Zeit}'',
Physik. Zeitschr. {\bf 10}, 104 (1909).
\bibitem{unruh}
W.\,G.~Unruh, ``{Experimental Black-Hole Evaporation?}'',
Phys.\ Rev.\ Lett.\ {\bf 46}, 1351 (1981).
\bibitem{visser} M.~Visser,
 ``{Acoustic black holes: horizons,
ergospheres, and Hawking radiation}'',
Class.\ Quantum Grav.\ {\bf 15}, 1767 (1998).
\bibitem{GrishaUniverse} G.\,E. Volovik,   
{\em The Universe in a Helium Droplet} (Oxford University Press, Oxford,
2003).
\bibitem{trautman} A. Trautman,
``{Comparison of Newtonian and
relativistic theories of spacetime}'', in: {\em Perspectives in
Geometry and Relativity} (Indiana University Press, Bloomington, 1966).
\bibitem{PG} P. Painlev\'e, ``{La m\'ecanique classique et la th\'eorie
de la relativit\'e}'',
C. R. Hebd. 
Acad. Sci. (Paris) {\bf 173}, 677 (1921);
A. Gullstrand,
``{Allgemeine L\"osung des statischen Eink\"orperproblems in der
Einsteinschen Gravitationstheorie}'',
Arkiv. Mat. Astron. Fys. {\bf 16}, 1 (1922).
\bibitem{Annalspaper} U.\,R. Fischer and M. Visser,
``{On the space-time curvature experienced by quasiparticle excitations
in the Painlev\'e-Gullstrand effective  geometry}'',
Ann. Phys. (N.Y.) {\bf 304}, 22 (2003).


\bibitem{madelung} E. Madelung,
``{Quantentheorie in hydrodynamischer Form}'',
Z. Phys. {\bf 40}, 322 (1927).
\bibitem{onsagerstathydro} L. Onsager, ``{Statistical Hydrodynamics}'',
Nuovo Cimento Suppl. {\bf 6}, 279 (1949).

\bibitem{BLV}
C.~Barcel\'o, S.~Liberati, and M.~Visser,
``{Analogue gravity from field theory normal modes?}'',
Class.\ Quantum\ Grav.\  {\bf 18}, 3595 (2001).

\bibitem{Jacobson}  T.\,A. Jacobson and G.\,E.  Volovik,
``Event horizons and ergoregions in $^3\!$He'',
Phys. Rev. D {\bf 58}, 064021 (1998).


\bibitem{dielectric}
W. Gordon, ``Zur Lichtfortpflanzung nach der Relativit\"atstheorie'',
Ann. Phys. (Leipzig) {\bf 72}, 421 (1923);
U. Leonhardt, ``{Space-time geometry of quantum
dielectrics}'', Phys. Rev. A {\bf 62}, 012111 (2000);
R. Sch\"utzhold, G. Plunien, and G. Soff,
``{Dielectric Black Hole Analogs}'',
Phys. Rev. Lett. {\bf 88}, 061101 (2002).

\bibitem{surfacewaves} R. Sch\"utzhold and W.\,G. Unruh,
``Gravity wave analogues of black holes'', Phys. Rev. D {\bf 66},
044019 (2002).


\bibitem{BirrellDavies} N.\,D. Birrell and P.\,C.\,W. Davies, 
{\em Quantum Fields in Curved Space} (Cambridge University Press,
1984).


\bibitem{BillRalf} W.\,G. Unruh and R. Sch\"utzhold,
``On slow light as a black hole analogue'', Phys. Rev. D {\bf 68}, 024008
(2003).

\bibitem{VisserPRL} M. Visser,
``{Hawking radiation without
black hole entropy}'',
Phys. Rev. Lett. {\bf 80}, 3436 (1998);
L.\,J.~Garay, J.\,R.~Anglin, J.\,I.~Cirac, and P.~Zoller,
``{Sonic Analog of Gravitational Black Holes in Bose-Einstein Condensates}'',
Phys.\ Rev.\ Lett.\ {\bf 85}, 4643 (2000).

\bibitem{GHEffect} P.\,O. Fedichev and U.\,R. Fischer,
``{Gibbons-Hawking Effect in the Sonic de Sitter Space-Time
of an Expanding Bose-Einstein-Condensed Gas}'',
Phys. Rev. Lett. {\bf 91}, 240407 (2003);
``{Observer dependence for the phonon content of the sound field living on
the effective curved space-time background of a Bose-Einstein condensate}'',
Phys. Rev. D {\bf 69}, 064021 (2004).

\bibitem{CPP}   P.\,O. Fedichev and U.\,R. Fischer,
 ``{``Cosmological''
quasiparticle production in harmonically trapped superfluid
gases}'',
Phys. Rev. A {\bf 69}, 033602 (2004).

\bibitem{BarceloCPP} C. Barcel\'o, S. Liberati, and M. Visser,
``{Probing semiclassical analog gravity in Bose-Einstein condensates
with widely tunable interactions}'',
Phys. Rev. A {\bf 68}, 053613 (2003).

\bibitem{Uhlmann}  U.\,R. Fischer and R. Sch\"utzhold,
``{Quantum simulation of cosmic
inflation in two-component Bose-Einstein condensates}'', 
Phys. Rev. A {\bf 70}, 063615 (2004); R. Sch\"utzhold,
``{Dynamical Zero-Temperature Phase
Transitions and Cosmic Inflation or Deflation}'',
Phys. Rev. Lett. {\bf 95}, 135703 (2005); 
M. Uhlmann, Y. Xu, and R. Sch\"utzhold,
``{Aspects of Cosmic Inflation in Expanding Bose-Einstein Condensates}'',
New J. Phys. {\bf 7}, 248.1-248.17 (2005).

\bibitem{BLVReview} C. Barcel\'o, S. Liberati, and M. Visser,
``{Analogue Gravity}'',  
Living Rev. Relativity {\bf 8},  12.1-12.113 (2005);
{\sf URL: http://www.livingreviews.org/lrr-2005-12}.


\bibitem{stone}
M.~Stone, ``{Acoustic energy and momentum in a moving medium}'',
Phys.\ Rev.\ E {\bf 62}, 1341 (2000);
%
``{Phonons and forces: Momentum {\em versus}
pseudomomentum in moving fluids}'',
 pp.~335 in M.~Novello, M.~Visser, and G.~Volovik (editors),
{\em Artificial Black Holes} (World Scientific, Singapore, 2002).

\bibitem{balbinot}
R.~Balbinot, S.~Fagnocchi, A.~Fabbri, and G.\,P.~Procopio,
``{Backreaction in acoustic black holes}'',
Phys. Rev. Lett. {\bf 94}, 161302 (2005);
R.~Balbinot, S.~Fagnocchi, and A.~Fabbri,
``{Quantum effects in acoustic black holes: The backreaction}'',
Phys. Rev. D {\bf 71}, 064019 (2005).

\bibitem{PRDpaper}
R. Sch\"utzhold, M. Uhlmann, Y. Xu, and U.\,R. Fischer,
``{Quantum backreaction in dilute Bose-Einstein condensates}'',
 Phys. Rev. D {\bf 72}, 105005 (2005).

\bibitem{Particle}
M.~Girardeau and R.~Arnowitt, ``{Theory of Many-Boson Systems: Pair Theory}'',
Phys.\ Rev.\ {\bf 113}, 755 (1959);
%
C.\,W.~Gardiner,
``{Particle-number-conserving Bogoliubov method which demonstrates
the validity of the time-dependent Gross-Pitaevski\v\i\/ equation for a
highly condensed Bose gas}'',
Phys.\ Rev.\ A {\bf 56}, 1414 (1997);
%
M.\,D.~Girardeau,
{\it ibid.}\ {\bf 58}, 775 (1998).

\bibitem{castin}
Y.~Castin and R.~Dum,
``{Low-temperature Bose-Einstein condensates in time-dependent traps:
Beyond the U(1) symmetry-breaking approach}'',
Phys.\ Rev.\ A {\bf 57}, 3008 (1998).

\bibitem{1/N}
E.\,H.~Lieb, R.~Seiringer, and J.~Yngvason,
``{Bosons in a trap: A rigorous derivation of the Gross-Pitaevski\v\i\/
energy functional}'', Phys.\ Rev.\ A {\bf 61}, 043602 (2000).

\bibitem{derivation}
The total particle number operator ${\hat N=\hat A^\dagger\hat A}$ and
the corresponding creation and annihilation operators satisfy
${[\hat A,\hat A^\dagger]=1}$ and
${[\hat\chi,\hat A{\hat N}^{-1/2}]=[\hat\zeta,\hat A{\hat N}^{-1/2}]=0}$,
i.e., the excitations $\hat\chi$ and $\hat\zeta$ are
particle-number-conserving, and thus the full
Hamiltonian can be written in terms of these operators,
cf.~\cite{Particle,castin}.
%
The mean-field ansatz in Eq.\,(\ref{mean--field}) can be motivated by
starting with $N$ free particles, $g=0$, in the same single-particle
state $\psi_{\rm c}$ with $\hat\zeta=0$ and subsequently switching on the
coupling $g>0$ by following the evolution in
Eqs.\,(\ref{GP})-(\ref{xi}) such that the corrections
$\hat\zeta={\cal O}(1/\sqrt{N})$ remain small \cite{meanfield}.

\bibitem{meanfield}
R. Sch\"utzhold, M. Uhlmann, Y. Xu, and U.\,R. Fischer,
``{Mean-field expansion in Bose-Einstein condensates with finite-range
interactions}'', 
Int. J. Mod. Phys. B {\bf 20,} 3555 (2006).    

\bibitem{GP}
E.\,P.~Gross,  ``{Structure of a Quantized
Vortex in Boson Systems}'',
Nuovo Cimento {\bf 20}, 454 (1961);
``{Hydrodynamics of a superfluid condensate}'',
J.\ Math.\ Phys.\ {\bf 4}, 195 (1963);
%
L.\,P.~Pitaevski\v\i\/, ``{Vortex lines in an imperfect Bose gas}'',
Sov.\ Phys.\ JETP {\bf 13}, 451 (1961).
%

\bibitem{Lee}
T.\,D.~Lee and C.\,N.~Yang,
%
``{Many-Body Problem in Quantum Mechanics and Quantum Statistical
Mechanics}'',
Phys.\ Rev.\ {\bf 105}, 1119 (1957);
%
T.\,D.~Lee, K.~Huang, and C.\,N.~Yang,
``{Eigenvalues and Eigenfunctions of a Bose System of Hard Spheres and
Its Low-Temperature Properties}'', 
{\it ibid.}\ {\bf 106}, 1135 (1957);
%
E.~Timmermans, P.~Tommasini, and K.~Huang,
``{Variational Thomas-Fermi theory of a nonuniform Bose condensate at
zero temperature}'', Phys.\ Rev.\ A {\bf 55}, 3645 (1997).

%

\bibitem{BdG}
N.\,N.~Bogoliubov, ``{On the Theory of Superfluidity}'',
J.\ Phys.\ (USSR) {\bf 11}, 23 (1947);
%
P.\,G.~de~Gennes, {\em Superconductivity of Metals and Alloys}
(W.\,A. Benjamin, New York, 1966).

\bibitem{temperature}
A.~Griffin,
``{Conserving and gapless approximations for an inhomogeneous Bose gas
at finite temperatures}'',
Phys.\ Rev.\ B {\bf 53}, 9341 (1996);
%
E.~Zaremba, A.~Griffin, and T.~Nikuni,
``{Two-fluid hydrodynamics for a trapped weakly interacting Bose gas}'', 
Phys.\ Rev.\ A {\bf 57}, 4695 (1998).

\bibitem{Yukalova} V.\,I. Yukalov and E.\,P. Yukalova, ``{Normal
and Anomalous Averages for Systems with Bose-Einstein Condensate}'',
Laser Phys. Lett. {\bf 2}, 506 (2005).

\bibitem{BLVBEC}
C.~Barcel\'o, S.~Liberati, and M.~Visser,
``{Analogue gravity from Bose-Einstein condensates}'',
Class.\ Quantum\ Grav.\  {\bf 18}, 1137 (2001).

\bibitem{Froehlich}
We note that the problem that the ``canonical''
commutator between density and phase
operators leads to fundamental inconsistencies was first
pointed out in the context of superfluid hydrodynamics by H. Fr\"ohlich,
``{A contradiction between quantum hydrodynamics
and the existence of particles}'',
Physica {\bf 34}, 47 (1967).

\bibitem{Castin} 
Y.~Castin, ``{Simple theoretical tools for low dimension Bose gases}'',
J.\ Phys.\ IV France {\bf 116}, 89 (2004).

\bibitem{Jaksch}  D. Jaksch, C. Bruder, J.\,I. Cirac, C.\,W. Gardiner,
and P. Zoller, ``Cold Bosonic Atoms in Optical Lattices'',
Phys. Rev. Lett. {\bf 81}, 3108 (1998).

\bibitem{1D} M.~Olshani\v\i,
``{Atomic Scattering in the Presence of an External Confinement
and a Gas of Impenetrable Bosons}'',
Phys.\ Rev.\ Lett.\ {\bf 81}, 938 (1998).

\bibitem{Goerlitz} A.~G\"orlitz {\em et al.},
``{Realization of Bose-Einstein condensates in lower dimensions}'',
Phys. Rev. Lett. \ {\bf 87}, 130402 (2001).

\bibitem{Uncertainty}
L.~Pitaevski\v\i\/ and S.~Stringari,
``{Uncertainty principle and off-diagonal long-range order in the
fractional quantum Hall effect}'',
Phys.\ Rev.\ B {\bf 47}, 10915 (1993).

\end{thebibliography}
\end{document}